# Tuning the period of femtosecond laser induced surface structures in steel: from angled incidence to quill writing

Yasser Fuentes-Edfuf[1,*] José A. Sánchez-Gil[2], Marina Garcia Pardo[1], Rosalía Serna[1], George D. Tsibidis[3], Vincenzo Giannini[2], Javier Solis[1], and Jan Siegel[1,†]

[1] *Laser Processing Group, Instituto de Optica (IO-CSIC), Consejo Superior de Investigaciones Científicas, Serrano 121, 28006 Madrid, Spain*

[2] *Instituto de Estructura de la Materia (IEM-CSIC), Consejo Superior de Investigaciones Científicas, Serrano 121, 28006 Madrid, Spain*

[3] *Institute of Electronic Structure and Laser (IESL), Foundation for Research and Technology (FORTH), N. Plastira 100, Vassilika Vouton, 70013, Heraklion, Crete, Greece.*

*Corresponding author: \* y.fuentes@csic.es, † j.siegel@csic.es*

## ABSTRACT

Exposure of metal surfaces to multiple ultrashort laser pulses under certain conditions leads to the formation of well-defined periodic surface structures. We show how the period of such structures in steel can be tuned over a wide range by controlling the complex interaction mechanisms triggered in the material. Amongst the different irradiation parameters that influence the properties of the induced structures, the angle of incidence of the laser beam occupies a prominent role. We present an experimental and theoretical investigation of this angle dependence in steel upon irradiation with laser pulses of 120 fs duration and 800 nm wavelength, while moving the sample at constant speed. Our findings can be grouped into two blocks. First, we observe the spatial coexistence of two different ripple periods at the steel surface, both featuring inverse scaling upon angle increase, which are related to forward and backward propagation of surface plasmon polaritons. To understand the underlying physical phenomena, we extend a recently developed model that takes into account quantitative properties of the surface roughness to the case of absorbing metals (large imaginary part of the dielectric function), and obtain an excellent match with the experimentally observed angle dependence. As second important finding, we observe a *quill writing* effect, also termed non-reciprocal writing, in form of a significant change of the ripple period upon reversing the sample movement direction. This remarkable phenomenon has been observed so far only inside dielectric materials and our work underlines its importance also in laser surface processing. We attribute the origin of symmetry breaking to the non-symmetric micro- and nanoscale roughness induced upon static multiple pulse irradiation, leading to non-symmetric modification of the wavevector of the coupled surface plasmon polariton.

## 1. INTRODUCTION

The interest in material processing with ultrafast lasers has risen enormously over the past decades due countless applications in technology and industry. Also for the particular case of metals, this strategy has allowed its functionalization in an unprecedented manner including, among others, the fine control of surface wettability, the significant reduction in friction and wear, or the fabrication of materials with tailored optical or electrochemical properties. These applications are only a glimpse of what can be achieved with this technology[1–6].

Many of the strategies used to fabricate functional structures with sub-micrometer feature sizes are based on an indirect structuring mechanism, rather than engraving the structure by direct writing. When a material surface is exposed to laser light under certain conditions in terms of pulse number, pulse duration, fluence, and light polarization, it is possible to fabricate self-organized surface structures, known as Laser Induced Periodic Surface Structures (LIPSS)[7]. LIPSS can be fabricated in metals[8–12], semiconductors[13–15], dielectrics[16–19], which has generated an enormous interest in the scientific community due in order to unravel the underlying mechanisms, as well as in industry because of the countless applications that are foreseen for these structures. Amongst the variety of different structure types that can be fabricated, the most common ones are the so-called ripples (parallel lines with a period near the laser wavelength usually perpendicular to its polarization), grooves (parallel lines with supra-wavelength period orientated parallel to the polarization), and spikes (disordered supra-wavelength cone-like structures).[13]

For the specific case of ripples, it is known that their period can be controlled up to a certain extent by changing the angle of incidence[15,20,21]. Assuming a simple scatter model for the incident laser light interfering with the light scattered at the surface, the following expression is obtained[22]:

$$\Lambda^{\pm} = \frac{\lambda}{1 \mp \sin\theta} \tag{1}$$

where $\theta$ is the angle of incidence with respect to the surface normal and $\lambda$ the laser wavelength. Notably, this model predicts two ripple periods $\Lambda^+$ and $\Lambda^-$ when irradiating at oblique incidence, which indeed has been observed experimentally in a variety of materials. Yet, the experimentally observed period values were found to obey eq. (1) only for the specific case of semiconductors[15,21]. For metals, the simple scattering model needs to be modified by taking into account the propagation of surface plasmon polaritons (SPP) at the surface, which yields[23,24]:

$$\Lambda^{\pm} = \frac{\lambda}{\text{Re}[\eta] \mp \sin\theta} \tag{2}$$

Where Re[η] is the real part of η, and $\boldsymbol{\eta} = [\epsilon_{air}\epsilon_{metal}/(\epsilon_{air}+\epsilon_{metal})]^{1/2}$ being the effective refractive index of the air-metal interface. $\epsilon_{air}$ and $\epsilon_{metal}$ are the dielectric functions of air and metal, respectively.

Although the vast majority of experimental studies is performed at normal incidence[25–32], some studies are performed at oblique incidence.[15,21,24,33–35] Interestingly, to the best of our knowledge, none of the experimentally determined ripple periods at elevated angles of incidence match the values predicted by either of the expressions (1) or (2). Numerous attempts to explain this discrepancy have been made, proposing modifications to the model. These include a transient change of $\epsilon_{metal}$ during laser irradiation,[25] the occurrence of complex hydrodynamics taking place after laser-induced melting,[13] as well as a modification of Re[η] by introducing an additional interface layer, whose optical properties are estimated by the Maxwell-Garnett theory of effective media [24,36]. Moreover, Ionin *et al.* introduced a modification to the model, based on the scattering of SPPs on surface relief features with a certain relief height, which predicts a corresponding shift of the SPPs' dispersion curve.[37] Their results obtained in silicon showed indeed a qualitative agreement of the model with their experimental data in terms of a trend that the period is reduced as a function of relief height, although no quantitative match was obtained. Very recently, we have introduced a related approach, based on the significant modification of the SPP wave vector by the presence of the surface roughness induced during multiple laser pulse irradiation, characterized by specific roughness parameters.[38] In that work, focusing on the ripple formation process in copper, we have introduced a model that takes into account the influence of the specific roughness properties on the SPP wave vector, and which is able to correctly describe the experimentally observed ripple periods.

In the present work, we perform a systematic study of the formation of ripples in steel upon irradiation with fs laser pulses, exploring the influence of a number of experimental parameters, with particular emphasis on the angle dependence and the writing direction. Moreover, we apply our recently developed model based on the modification of the SPP wave vector by the surface roughness to the experimental data.

## 2. MATERIALS AND METHODS

The samples used were made of commercially available steel 1.7131 (16MnCr5), mirror-polished down to a root mean square roughness $R_{RMS}$ < 20 nm. Laser irradiation was performed with an amplified, Ti:sapphire laser providing 120 fs pulses at 800 nm wavelength and 100 Hz repetition rate. The p-polarized beam was sent through a half-wave plate and a polarizing beam cube to select the incident pulse energy, followed by a circular aperture with a diameter of $\emptyset_1 = 3.5\ mm$ (or in certain occasions $\emptyset_2 = 5\ mm$) and a spherical lens of focal length $f = 150\ mm$ to focus the beam onto the sample surface. The Gaussian beam diameter (spot size) at $1/e^2$, determined at the sample surface using a method proposed by Liu[39], was $d_{x1} = 59$ μm or $d_{x2} = 44\ \mu m$ (for the apertures $\emptyset_1$ or $\emptyset_2$, respectively). The sample was mounted on a rotation stage in order to select the angle of incidence θ. Most experiments were performed by exposing the sample to laser pulses at constant pulse repetition rate (100 Hz) while moving the sample at a specific velocity. The effective pulse number per unit area for a given spot size and sample speed $v$ was defined as $N_{eff} = d_x(\theta) \cdot 100\frac{Hz}{v}$, with typical speeds in the range of 0.1 mm/s - 2 mm/s. Two scan directions where used. The initial scan direction was from right-to-left, defined hereafter as Forward Movement (FM), and the same set of experiments was repeated with the inverted scan direction left-to-right defined hereafter as Backward Movement (BM). For the specific results shown in Figure 6, relatively large areas needed to be fabricated which was achieved by writing an array of parallel lines in FM mode with a line spacing of s = 65 μm, which was determined empirically as the optimum spacing for fabricating homogeneous ripples. In this case, aperture $\emptyset_1$ was used, leading to a small pulse overlap between lines, which was taken into account as $N_{eff\_2D} = \frac{\pi}{4}d_{x1}(52º) \cdot 100\ Hz/(v \cdot s)$.

Since for oblique incidence the spot size is elliptic and larger than for normal incidence, the incident pulse energy needed to be increased in order to obtain the same $F$ value. To this end, the long spot axis was determined as $d_x(\theta) = d_y/\cos(\theta)$, with the short axis being constant. It is important noticing that the absorbed fraction of the laser pulse energy is also angle dependent due to the corresponding dependence of the Fresnel reflection coefficient. This was taken into account by formulating an effective fluence $F_{eff} = (1 - R(\theta)) \cdot 8E \cdot \cos(\theta)/(\pi \cdot d_y^2)$. The angle dependent reflectivity and dielectric functions of the material were measured via spectral ellipsometry. All experiments were performed in ambient air. The surface morphology was characterized with scanning electron microscopy (SEM) and optical microscopy (OM). Ripple periods were determined via performing Fast Four Transformation (FFT) analysis of the corresponding microscope images and the error extracted corresponds to two times the standard deviation of a Gaussian fit to the first order of the FFT, as explained in more detail in the appendix. The surface topography of the structures was measured with an atomic force microscope (AFM) operating in tapping mode and the surface roughness values were determined via statistical analysis using the freeware Gwyddion.

## 3. RESULTS AND DISCUSSION

### 3.1. Angle dependence of the ripple period

The angle dependence of the ripple period was investigated for a constant pulse number $N_{eff} = 100$ and effective fluence $F_{eff} = 1.0\ J/cm^2$ and selected results are shown in form of SEM images in *Figure 1*. For $\theta = 0°$, well-pronounced ripples with a period $\Lambda = 0.7 \pm 0.2\ \mu m$ and orthogonal orientation with respect to the laser polarization are formed, decorated with small droplets of nanometer size. As predicted by eq. (2), for higher angle values this single period splits into a long period $\Lambda^+$ and a short period $\Lambda^-$, which is especially evident at $\theta = 40°$ and $\theta = 52°$. A look to the higher magnification images reveals a significant decrease of $\Lambda^-$ compared to $\Lambda(0°)$.

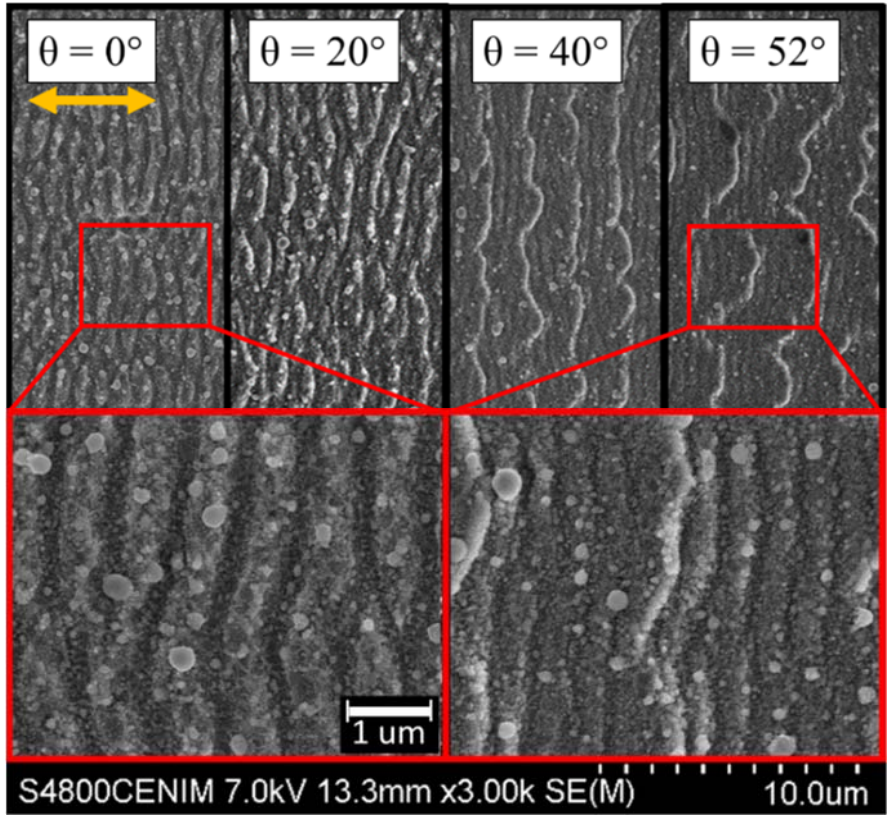

*Figure 1: SEM images of ripple structures formed in steel at $F_{eff}$ = 1.0 ± 0.1 J/cm² and $N_{eff}$ = 100. Top row: Structures obtained at different angles of incidence **θ**, indicated in each frame. The laser polarization orientation is indicated by the double-headed arrow. Bottom row: Higher magnification images of the areas marked in the corresponding images above (**θ = 0°** (left) and **θ = 52°** (right)).*

Figure 2 displays the evolution of the experimentally determined ripple periods $\Lambda^+$ and $\Lambda^-$ as a function of angle $\theta = [0, 10, 20, 30, 40, 52, 60]°$. The figure also includes, in form of dashed lines, the periods $\Lambda^+_{SPP}$ and $\Lambda^-_{SPP}$ predicted by the simple SPP model in eq. (2), using Re[η] = 1.01 calculated using the dielectric functions for steel and air. An increasing trend for $\Lambda^+$ and decreasing trend for $\Lambda^-$ can be observed, as expected for the angle dependence. Whereas an apparently good quantitative agreement is obtained for $\Lambda^-$, the absolute values of $\Lambda^+$ at higher angles are significantly below those predicted by the model (cf. eq. (2)). As mentioned in the introduction, this deviation is generally observed in metals and numerous models have been proposed to correct for it,[13,24,25,36] including our recently introduced model based on a roughness-mediated change of the SPP wave vector.[40]

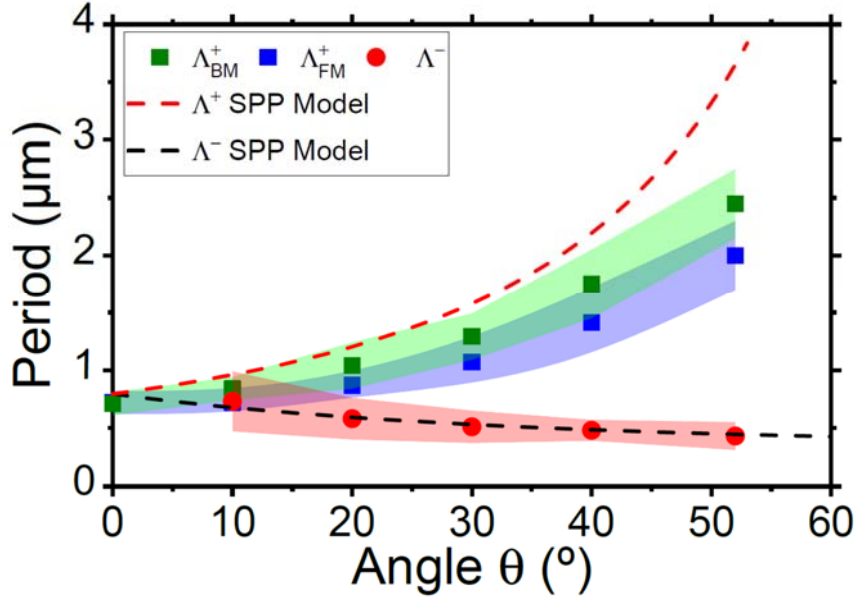

*Figure 2: Ripple periods as a function of the angle of laser incidence when moving the sample. The effective pulse number and fluence are respectively* $N_{eff} = 100$ *and* $F_{eff} = 1.0 \pm 0.1 \, J/cm^2$. *Symbols correspond to experimental data for* $\Lambda^+$ (blue and green squares) *and* $\Lambda^-$ (red circles) *and shaded areas indicate the experimental error in ripple determination (see experimental section). The two values per angle for* $\Lambda^+$ *correspond to the different values obtained when moving the sample along different directions, denominated* $\Lambda^+_{BM}$ *and* $\Lambda^+_{FM}$. *Dashed curves correspond to the values* $\Lambda^+_{SPP}$ *and* $\Lambda^-_{SPP}$ *calculated with the simple SPP model using eq.* (2) *(black: short period, red: long period).*

### *3.2. Quill writing: Non-reciprocal ultrafast laser writing*

The reader will have noticed that in Figure *2* two values per angle are plotted for the $\Lambda^+$, namely $\Lambda^+_{BM}$ and $\Lambda^+_{FM}$, corresponding respectively to the backward and forward movement direction of the sample upon scanning. This reflects the experimental fact that significant differences in the ripple period are observed when reversing the movement direction of the sample. As can be seen in the figure, this effect is observed only for oblique incidence and increases with angle. An illustrative example is shown in Figure 3. Irradiations were performed at $\theta = 52°, F_{eff} = 1.0 \pm 0.1 \, J/cm^2$ with a sample movement velocity of v = 0.104 mm/s, which corresponds to $N_{eff} = 100$ for the pulse repetition rate used (100 Hz). The initial scan direction from left-to-right (Figure 3a), defined hereafter as Backward Movement (BM), was inverted to right-to-left (Figure 3b), defined as Forward Movement (FM). In both cases, ripples with similar appearance and two different periods were formed, as can be seen in the SEM images (cf. Figure 3). However, performing Fourier transformations on the SEM images reveals a systematic shift of the position of the first order between FM and BM for oblique angles of incidence, as can be seen in Fig. 2.

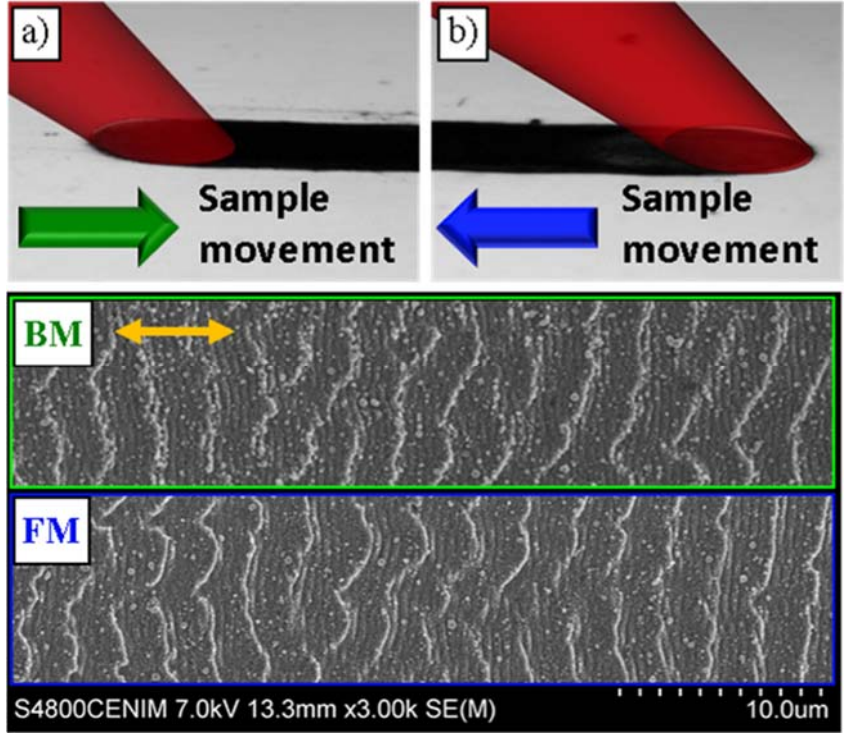

*Figure 3: Top row: Schemes of the irradiation configuration at $100\ Hz$ repetition rate while moving the sample in different directions. Middle and Bottom: BM and FM correspond to SEM images of structures written while moving the sample either to the right (BM) or to the left (FM) under otherwise identical conditions ($N_{eff} = 100$, $F = 1.0 \pm 0.1\ J/cm^2$, $\theta = 52°$). The two-head arrow represents the polarization direction.*

The resulting period values obtained for $\theta = 52°$ are $\Lambda^+_{BM} = 2.4 \pm 0.3$ µm and $\Lambda^+_{FM} = 2.0 \pm 0.3$ µm. One may argue that the difference might lie within the error in the determination of the period via FFT, and thus attribute the observed deviation to a mere statistical fluctuation. However, this possibility can be ruled out since we have repeated the same experiment many times, as well as explored different $N_{eff}$ values, systematically observing $\Lambda^+_{BM} > \Lambda^+_{FM}$.

Such behavior, when the properties of a written surface structure depend on the writing direction, is unusual. In the field of laser writing, a similar direction dependence has been observed already in subsurface writing of optical waveguides inside glasses, first reported by Kazansky et al.[41,42] The authors have coined the term "non-reciprocal writing" or "quill writing" in analogy to the different appearance of letters when writing with a quill and ink from left-to-right or right-to-left. The authors attributed this effect in their specific case to a pulse front tilt of the focused laser beam, which is often present in experimental systems due to minor misalignments of optical components. This tilt in the intensity distribution in the front of the incident pulse causes a certain asymmetry in plasma distribution upon focusing the pulse inside the material, even at normal incidence.[41] Non-reciprocal writing has also been observed upon surface processing, although it has not been explicitly named using the terms mentioned above. Ionin *et al.* employed fs laser beam shaping to generate an asymmetrical spatial fluence distribution, which allowed the fabrication of different types of nano- and microstructures at normal incidence, just by changing the scanning direction.[43] Likewise, non-reciprocal writing of self-organized surface structures in silicon has been reported recently without using the terminology coined by the Kazansky group.[15]

While the results reported in the present work are fully consistent with the *quill writing* effect observed by Kazansky et al.[41] and Ionin et al.[43] (the period depends on the scan direction), we have found that in our case the underlying origin is different from that reported previously. When processing the surface of a strongly absorbing material at normal incidence, a possible pulse front tilt has only a minimal effect for a symmetric plasma distribution has a thickness of only a few tens of nanometers.

Likewise, using a radially symmetric Gaussian intensity profile, as in our case, does not provide the required asymmetry for non-reciprocal surface writing. The same holds for a situation in which the sample is moved at constant velocity, since the laser-induced roughness distribution, which increases with pulse number, is symmetric along the axis of the electric field vector and a 180°-change of scan direction leaves the system invariant. However, the situation changes when processing at oblique incidence, since the laser-induced roughness distribution becomes asymmetric, as we will show next.

Figure 4 shows a comparison of two areas exposed to $N = 100$ laser pulses at $F_{eff} = 1.0 \pm 0.1\, J/cm^2$, at two different angles of incidence, without scanning the sample. For normal incidence ($\theta = 0°$), the morphology distribution is symmetric with respect to both, the horizontal and vertical axes, as can be seen in the main frame and the magnified regions. This suggests that upon sample scanning along the horizontal axis ripples with the same period will be obtained, independently from the direction. However, for $\theta = 52°$, symmetry is only observed along with respect to the horizontal axis but is broken with respect to the vertical axis. The pronounced differences in morphology between the left and the right part of the laser-modified areas can be clearly seen in the magnified regions. While the left part of the spot is composed essentially of vertical ripples of type $\Lambda^-$ at the outmost region and horizontally aligned ripples of different origin (not discussed further) towards the spot center, the right part of the spot features a mirrored morphology but is superimposed with vertical ripples of type $\Lambda^+$. This asymmetry in morphology is the origin of the *quill writing* effect upon ripple fabrication when moving the sample, leading to the different periods shown in Figure 3.

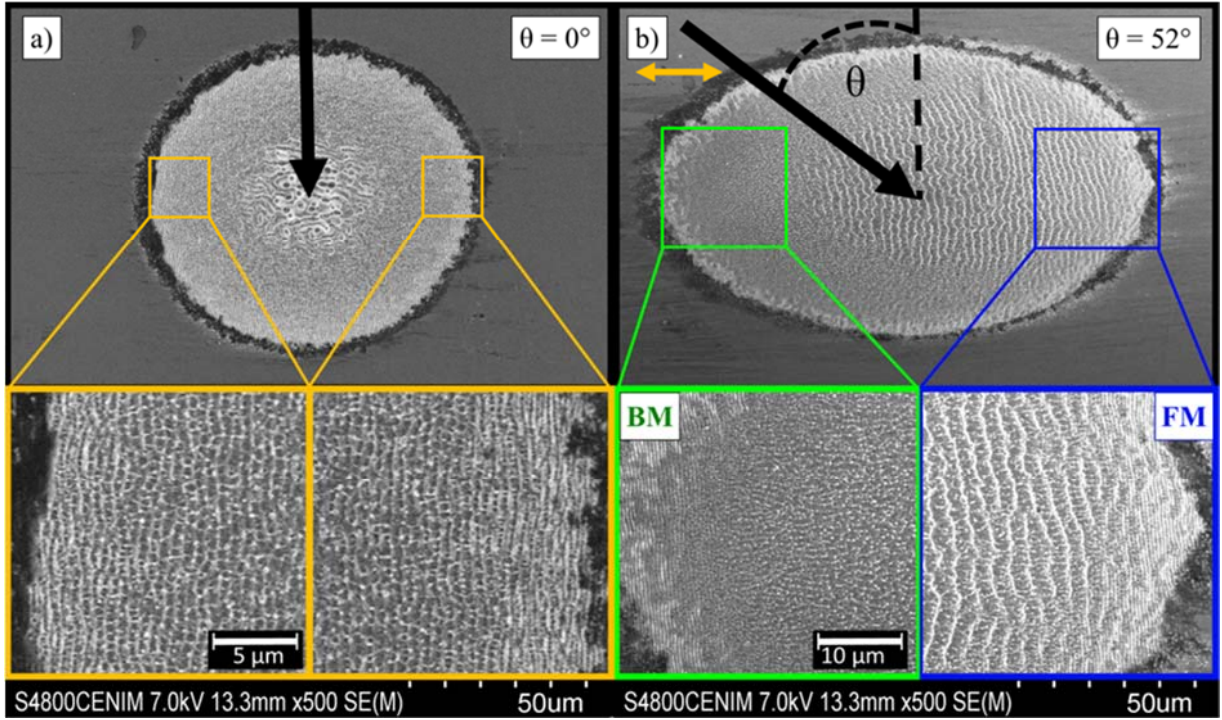

*Figure 4: SEM images of static irradiation at two different angles of incidence with $\bm{N = 100}$, $\bm{F_{eff}} = \bm{1.0 \pm 0.1\, J/cm^2}$. (a) $\bm{\theta = 0°}$, (b) $\bm{\theta = 52°}$. The bottom images are magnified areas from the left and right part of the irradiated area. The orange two-head arrow represents the polarization. The arrows indicate the direction/angle of incidence of the laser pulses.*

It is well known that the ripple period can be influenced by the pulse number, even for experiments performed at normal incidence.[44] In order to investigate the influence of the pulse number on the ripple period in presence of the *quill writing* effect upon dynamic irradiation, we have performed a corresponding experiment at oblique incidence and for both scan directions. The results are plotted in Figure 5. At the pulse number threshold for ripple formation ($N_{eff} = 20, v = 0.479\, mm/s$), a single period value is obtained, $\Lambda^+ = 2.2$ μm, with $\Lambda^-$ ripples being absent. As the pulse number increases to $N_{eff}$ = 40, $\Lambda^-$ ripples appear with a period of about 0.45 µm. In addition, the $\Lambda^+$ period splits into two, a slightly larger one for backward movement $\Lambda^+_{BM}$ and a smaller one for forward movement $\Lambda^+_{FM}$. The difference between both values increases continuously, amounting to a maximum of $\Delta\Lambda^+ = 0.6$ µm at $N_{eff} = 200$. This result shows that the *quill writing* effect is enhanced for higher pulse numbers and reminds of the *quill writing* effect in bulk glasses, in which this effect increases as a function of front pulse tilt.[45] Note that in Figure 5 both $\Lambda^+$ periods always stay well below the value predicted by the simple plasmonic model (red dashed line), demonstrating that the overall deviation from the model is not a pulse number effect. Concerning $\Lambda^-$, no evidence of period splitting can be observed and the experimental value matches well the predictions from the simple plasmonic model (black dashed line). A possible explanation for this apparent match with the simple model is the fact that a possible deviation would always be a relative deviation and therefore not show up for the small period values in that case. For instance, for $\Lambda^+$ the deviation from the model is $\Delta\Lambda^+$ = (3.8 µm – 2.2 µm)/ 3.8 µm = 1.6 µm = - 42%. Applying this percentage to the short period, we would expect $\Lambda^-$ = 0.45 µm + 42% = 0.64 µm. This expected value lies just within the error bar of the experimental data ($\Lambda^-$ = 0.45 µm ± 0.2 µm). We can therefore not unambiguously state that the simple SPP model is valid for the short period.

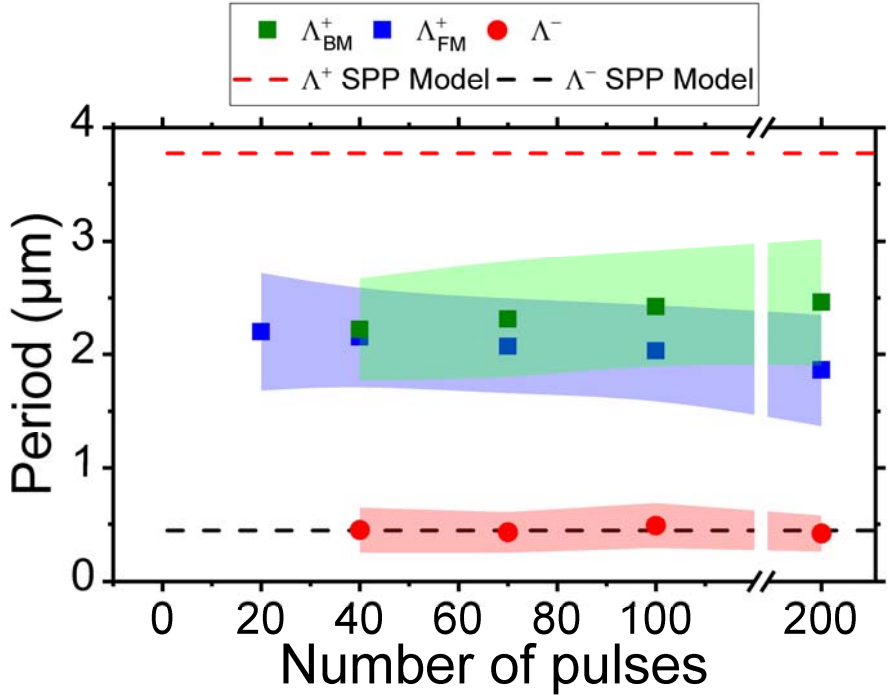

*Figure 5: Plot of the experimentally measured ripple periods $\Lambda^+_{BM}$, $\Lambda^+_{FM}$ and $\Lambda^-$ at a fixed angle of incidence $\theta = 52°$ as a function of pulse number $N_{eff}$ for a moving sample. The irradiations where performed at $F_{eff} = 1.0 \pm 0.1\, J/cm^2$. Green squares (backward movement) and blue squares (forward movement) correspond to long periods. For the short period (red circles) a single value is obtained for both scan directions. The shaded regions mark the experimental error in the determination of the period from the FFT. Dashed lines represent the periods $\Lambda^+_{SPP}$ (red) and $\Lambda^-_{SPP}$ (black) predicted by the simple SPP Model.*

### 3.3. Roughness-mediated SPP propagation

As we have recently shown for LIPSS formation in Cu,[40] the surface roughness induced by the laser pulses plays a major role in the ripple period obtained, namely, through a modification of the wave vector of the surface plasmon polariton. In order to investigate the possible presence of this effect in steel, we have performed AFM measurements recorded with a step size $d = 20\, nm$ of regions processed at $\theta = 52°$ for different $N_{eff}$ values. Fig. 6 (top) shows as an example AFM profiles along the direction of laser polarization for $N_{eff}$ = 10 and $N_{eff}$ = 20, corresponding respectively to below and above the ripple formation threshold. While an irregular topography with a superimposed nano-roughness can be observed in the former case, a periodic modulation with less roughness has emerged in the latter. We have analyzed the AFM maps for all $N_{eff}$ values and extracted the corresponding RMS height $\delta_{nano}$ of the nano-roughness for those maps featuring no periodic modulation, whereas for $N_{eff} \geq 20$, the modulation depth (peak-valley) of the ripples is displayed. The results displayed in Figure 6(bottom) show that the first pulses increase the surface roughness gradually until the ripple formation threshold (open squares), leading to a considerable modulation depth (full squares). This figure also displays the evolution of the sample reflectivity for 800 nm p-polarized light incident at $\theta = 52°$ as a function of pulse number, measured by ellipsometry on the laser-irradiated areas. The purpose of this figure is to analyze a possible change in energy coupling of the p-polarized laser pulse as the pulse number increases. A very strong decrease from an initial high reflectivity of R = 0.45 before irradiation down to R = 0.12 occurs within the first 20 pulses, there almost stabilizing upon a further increase in pulse number. This behavior can be interpreted as a strong increase of energy coupling triggered by the first pulses, consistent with the observed increase of roughness that causes light scattering and coupling of SPPs.

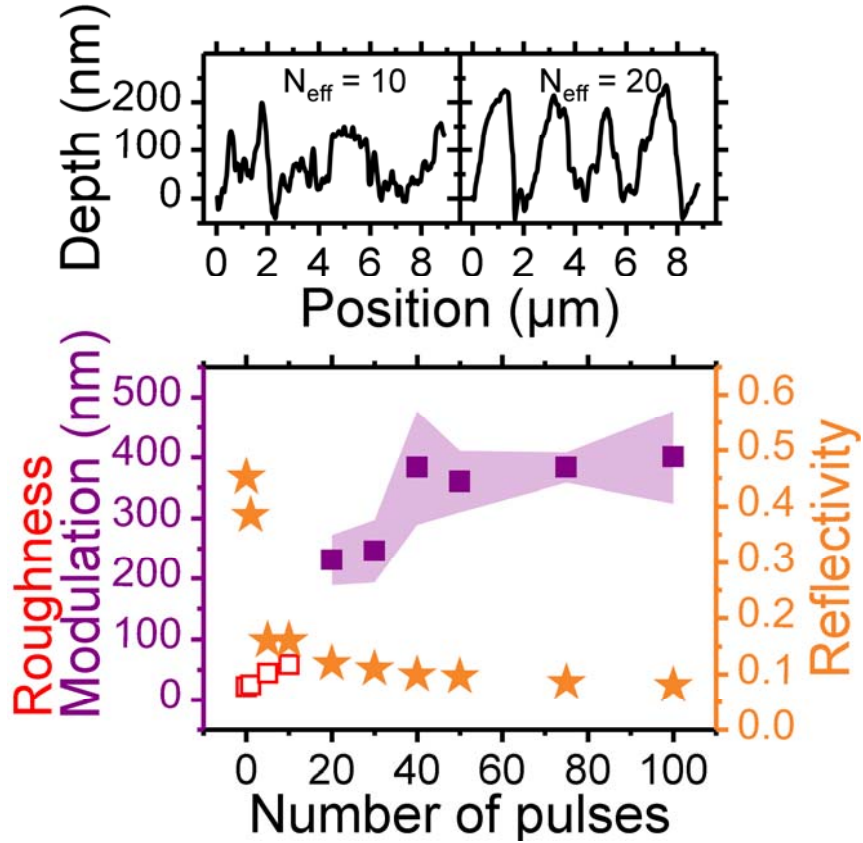

*Figure 6: (top) Depth profiles perpendicular to the ripple orientation extracted from AFM measurements of areas irradiated at $F_{eff} = 1.0 \pm 0.1 \, J/cm^2$ and $\theta = 52°$ incidence (see experimental section) for two different pulse numbers $N_{eff}$. (bottom) Evolution of the surface roughness/modulation depth (open and full square symbols, respectively) and surface reflectivity at 800 nm light illumination (star symbols) at $\theta = 52°$ incidence as a function of pulse number. The shaded region for the modulation depth indicates the experimental error in the determination of the period from the FFT, whereas the error bars for the other quantities is smaller than the corresponding symbol size. For $N_{eff} = [1, 10]$, no ripples are formed and the plotted data corresponds to the RMS height $\delta_{nano}$ of the nano-roughness, whereas for higher values, the modulation depth of the ripples is displayed.*

With this detailed information on the emergence of surface roughness with increasing number of pulses, we calculate the roughness-induced modification of the SPP wavevector. Following Refs. [40,46,47], it can be expressed as:

$$\Delta k_{SPP} = \frac{\delta^2 \sigma^2}{2} \frac{|\epsilon_r|^{1/2}}{(\epsilon_r + 1)^2} exp\left[-\left(k_{SPP}^{(0)}\right)^2/4\right] \\ \times \int_0^\infty kdk \frac{\alpha - \epsilon\alpha_0}{k^2 - \left(k_{SPP}^{(0)}\right)^2} exp\left[-\frac{k^2\sigma^2}{4}\right] F\left(k, k_{SPP}^{(0)}, \sigma\right), \quad (3)$$

with

$\epsilon_{metal} = \epsilon_r + i\epsilon_i$, $|\epsilon_r| \gg |\epsilon_i|$, and $\alpha_0 = \left[k^2 - \left(\frac{\omega}{c}\right)^2\right]^{1/2}$, $\alpha = \left[k^2 - \varepsilon\left(\frac{\omega}{c}\right)^2\right]^{1/2}$. This modifies the effective refractive index of the air-metal interface that has to be included in Eq. (2) in this manner:

$$Re(\eta') = Re(\eta) + \Delta k_{SPP} \cdot \lambda/(2\pi). \quad (4)$$

Here, we consider surface roughness as a Gaussian-correlated, Gaussian statistics random process, characterized by the RMS height deviation $\delta$ and by its correlation length $\sigma$, both described in more detail in Ref. [40]. Incidentally, it should be recalled that the expression (3) has been obtained through surface roughness perturbation theory[46,48], fully justified for the deeply subwavelength surface roughness dealt with here ($\delta_{nano}, \sigma_{nano} \ll \lambda$), but also assuming that the imaginary part of the dielectric function is much smaller than the real part,[46] $|\epsilon_r| \gg |\epsilon_i|$, which does not hold for steel. Since we are interested only in the real part of the SPP wavevector, a common approach to do so formally[49,50] is to replace the actual dielectric function of steel by an effective one $\epsilon'_{metal}$ with negligible absorption, $\epsilon'_i = 0$, and a real part $\epsilon'_r = \epsilon'_{metal}$ that yields the same SPP wavevector at a planar steel surface: $Re(\eta) = [\epsilon_{air}\epsilon'_{metal}/(\epsilon_{air}+\epsilon'_{metal})]^{1/2}$.

Calculations have been performed for different angles of incidence and the results are shown in Fig. 7 together with the experimental data (already shown in Figure 2). We have used the experimentally determined roughness parameters ($\delta$ = 59 nm and $\sigma$ = 390 nm, obtained under FM conditions) just before ripple formation (i.e. at $N_{eff}$ = 10, see Fig. 6) for the calculation of the predicted $\Lambda_{FM}^+$ and $\Lambda_{FM}^-$ periods. A very satisfactory fit is obtained for both periods (Fig. 6, solid blue line), which strongly supports the validity of the roughness-induced SPP model to correctly describe LIPSS formation not only for metals with negligible absorption (small imaginary part of the dielectric function) such as Cu,[40] but also for absorbing metals such as steel. A slightly better fit (not shown) can be obtained, when leaving one of the roughness parameters as fit parameters, yielding $\delta$ = 65 nm at fixed $\sigma$ = 390 nm. This could indicate that using the experimentally measured surface roughness at $N_{eff}$ = 10 underestimates the threshold pulse number at which the random roughness transforms into a periodic structure. It should be reminded here,[40] that the criterion for determining the appropriate $N_{eff,thres}$-value for specific experimental conditions is using the highest $N_{eff}$-value that leads to a random nano-roughness, not featuring periodic structures, which in our case lies within the range $N_{eff}$ = [10 - 20].

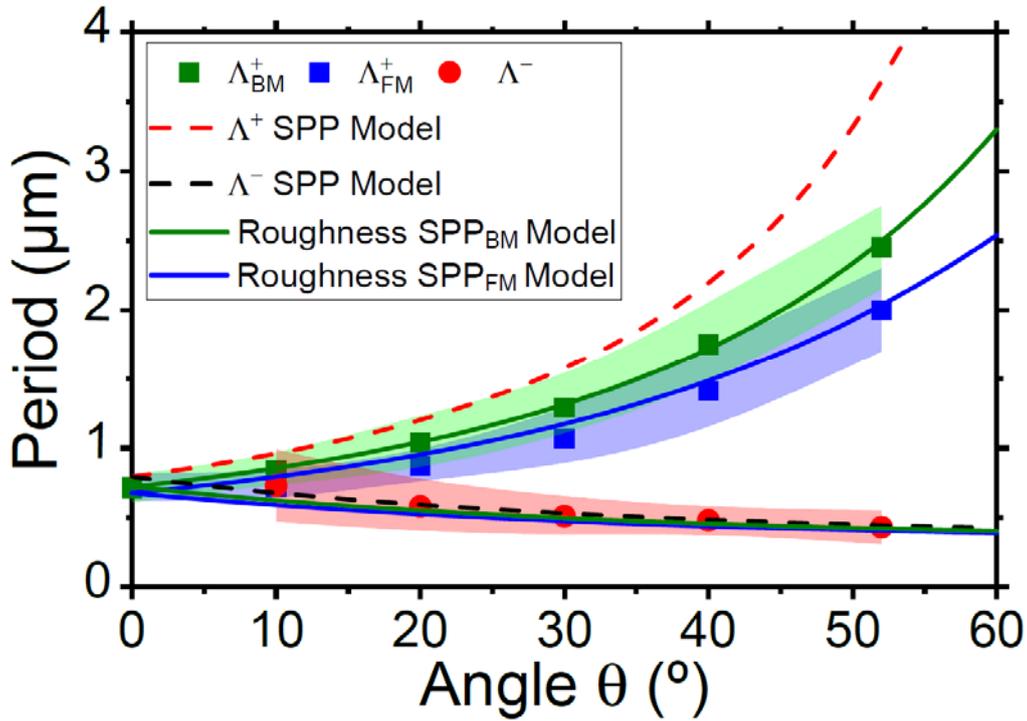

*Figure 7: Ripple periods as a function of incidence angle: Comparison of experimental data (symbols, conditions as for Fig. 2) with the simple SPP model using eq. (2) (black: short period, red: long period) (dashed lines) and the new model based on a roughness-mediated change of the SPP wave vector (solid lines). The roughness parameters used in the latter model were δ = 59 nm and σ = 390 nm for FM (determined from AFM measurements) and δ = 45 nm and σ = 390 nm for BM.*

Another possible reason for a deviation from the calculated curve from the experimental data points is the fact that the experimental roughness parameters have been determined for a single angle only (θ = 52°) and applied in the model to all angles. This very crude simplification ignores the possibility that different angles of incidence might lead to different surface roughness, even when keeping $N_{eff}$ and $F_{eff}$ constant.

The predicted $\Lambda^+_{BM}$ and $\Lambda^-_{BM}$ periods have been calculated by leaving the RMS height as a fit parameter and using the same correlation length value as determined experimentally for $\Lambda^+_{FM}$. Also here, a very good match of the model is observed, yielding a slightly lower RMS height for ripples obtained via backward movement (δ = 45 nm and σ = 390 nm). Variations in the short ripple periods are consistent with the model, but too small to be properly discerned.

**CONCLUSIONS:**

We have investigated the influence of the angle of incidence on the formation of laser induced periodic surface structures in steel formed upon irradiation with ultrashort laser pulses while scanning the sample at different speeds. At an effective fluence of 1 J/cm², low spatial frequency periodic ripples decorated with nanoparticles are formed for effective pulse numbers of 20 or higher. The ripple period formed at normal incidence splits into two coexisting periods for oblique incidence, one that increases which angle and one that decreases. This inverse scaling behavior can be qualitatively understood in terms of forward and backward propagation of surface plasmon polaritons, which scales accordingly with the angle of incidence. For a quantitative description of the experimental period values we have extended our recently introduced model to absorbing metals (small imaginary part of the dielectric function), such as steel. The model is based on a surface roughness-induced modification of the SPP wave vector and the results demonstrate that it is able to successfully describe the experimental data.

We also report here, for the first time to the best of our knowledge, a subtle change in the ripple period formed at oblique incidence when reversing the movement direction of the sample. We identify this phenomenon as the surface-equivalent process to the *quill writing* effect observed during subsurface writing of optical waveguides inside glasses, first reported by the Kazansky group. We explain the underlying mechanism that triggers the *quill writing* effect at the surface as a result of an asymmetry of the laser-induced roughness distribution at oblique incidence upon multiple pulse irradiation. This asymmetry leads to different values of the wave vector of the coupled surface plasmon polariton at the two sides of the laser spot, and therefore different ripples periods when scanning towards the direction of one side or the other.

**ACKNOWLEDGEMENTS:**

JSi, JSo and YF acknowledge the Spanish Ministry of Science, Innovation and Universities for financial support through research grant UDiSON (TEC2017-82464-R), the European Commission for grant LiNaBioFluid (FETOPEN 665337) and Consejo Superior de Investigaciones Científicas for the project INTRAMURALES 201850E057. VG and JASG acknowledge the Spanish Ministerio de Economia y Competitividad for financial support through the grants LENSBEAM (FIS2015-69295-C3-2-P) and NANOTOPO (FIS2017-91413-EXP), and also Consejo Superior de Investigaciones Científicas (INTRAMURALES 201750I039). G.D.T. acknowledges financial support from Nanoscience Foundries and Fine Analysis (NFFA)–Europe H2020-INFRAIA-2014-2015 (under Grant agreement No 654360and COST Action TUMIEE (supported by COST-European Cooperation in Science and Technology).

**APPENDIX:**

In order to determine the ripple periods we have employed a systematic procedure to calculate both the mean value of the period distribution, as well as distribution width. Two-dimensional Fast Fourier Transformations of SEM images and optical micrographs at moderate magnifications provide information about the presence of periodic structures. This two-dimensional analysis in the Fourier space was chosen, since it allows analyzing entire images instead of measuring individual periods in the original images and performing statistics, thus yielding more representative ripple values. Fig. 8(a) shows a representative SEM image together with its 2D-FFT. The two, relatively narrow and elongated intensity distributions in the FFT image correspond to the spatial frequencies of the coarse ripples.

We then perform radial integration over the area containing the one of these distributions (yellow dashed lines at the figure inset) and plot the result in the spatial domain (see *Figure 8* b)) and perform a Gaussian fit over the distribution (red dashed line). The peak value of the Gaussian fit corresponds to the mean value of the ripple period. We take two standard deviations from the Gaussian fit as the characteristic width of the distribution of periods.

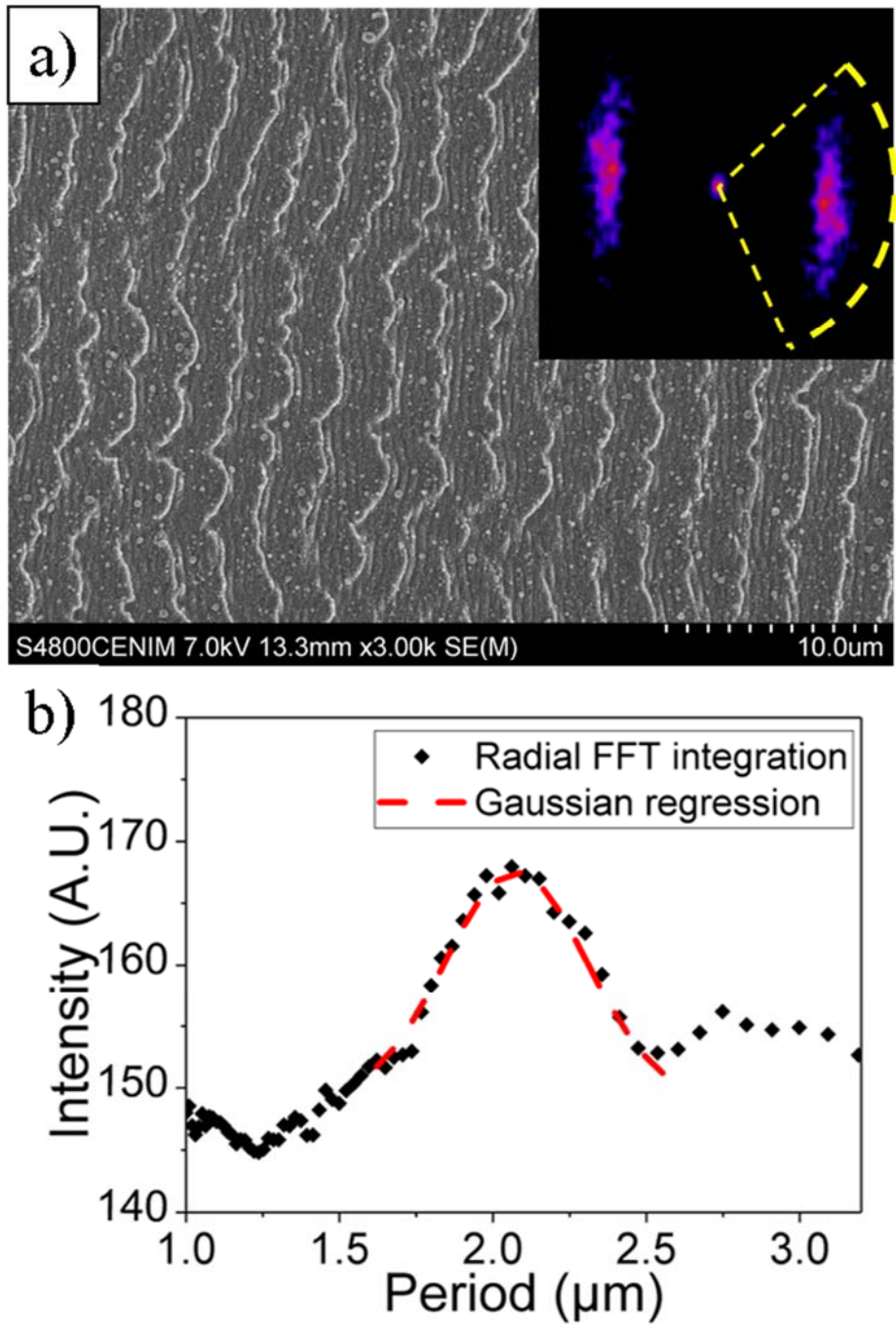

*Figure 8: a) SEM images of ripple structures at the steel surface. Inset: FFT of the image b) Plot of the distribution of ripple periods (black dots) obtained by radial integration of the FFT image intensity in the area delimited by yellow dashed line in a). The calculated Gaussian fit is represented by a red dashed line.*